# Al$_3$Sc thin films for advanced interconnect applications

Jean-Philippe Soulié,[a,*] Kiroubanand Sankaran,[a] Valeria Founta,[b,c,a] Karl Opsomer,[a] Christophe Detavernier,[d] Joris Van de Vondel,[c] Geoffrey Pourtois,[a,e] Zsolt Tőkei,[a] Johan Swerts,[a] and Christoph Adelmann,[a]

[a] Imec, 3001 Leuven, Belgium

[b] KU Leuven, Department of Materials Engineering, SCALINT, 3001 Leuven, Belgium

[c] Department of Physics and Astronomy, Quantum Solid-State Physics, 3001 Leuven, Belgium

[d] Ghent University, Department of Solid State Sciences, COCOON, 9000 Gent, Belgium

[e] Department of Chemistry, Plasmant Research Group, University of Antwerp, 2610 Wilrijk, Belgium

* Corresponding author. Email: jean-philippe.soulie@imec.be

**ABSTRACT**

Al$_x$Sc$_{1-x}$ thin films have been studied with compositions around Al$_3$Sc ($x$ = 0.75) for potential interconnect metallization applications. As-deposited 25 nm thick films were x-ray amorphous but crystallized at 190°C, followed by recrystallization at 440°C. After annealing at 500°C, 24 nm thick stoichiometric Al$_3$Sc showed a resistivity of 12.6 μΩcm, limited by a combination of grain boundary and point defect (disorder) scattering. Together with *ab initio* calculations that found a mean free path of the charge carriers of 7 nm for stoichiometric Al$_3$Sc, these results indicate that Al$_3$Sc bears promise for future interconnect metallization schemes. Challenges remain in minimizing the formation of secondary phases as well as in the control of the non-stoichiometric surface oxidation and interfacial reactions with underlying dielectrics.

**INTRODUCTION**

Today, Cu-based interconnects are progressively limiting the performance of large-scale integrated microelectronic circuits due to strongly increasing line and via resistances as well as reduced reliability.[1-4] The resistance increase originates from the reduced cross-sectional area of scaled interconnects in combination with the rapidly growing Cu resistivity at nanoscale dimensions. The resistivity increase is due to a long mean free path (MFP) of the charge carriers in Cu (of around 40 nm) that renders Cu very sensitive to finite size effects.[2,4-6] Moreover, Cu metallization requires barrier and liner layers to ensure reliable interconnect operation, which further reduces the Cu cross-sectional area without contributing (much) to the line or via conductance, therefore further aggravating the resistance issue.

For this reason, alternatives to Cu metallization have been intensively researched in recent years.[1,4,7,8] A particular focus has been on elemental metals with short MFPs and high melting temperatures, such as Co,[9,10] which has been introduced into commercial circuits recently,[11] Pt-group metals,[12] especially Ru,[13-15] or Mo.[16] Lately, the search for alternative metals has been broadened to include also binary and ternary intermetallics.[17-19] Among binary compound metals, only ordered intermetallics exhibit low bulk resistivities, rendering them potentially interesting for interconnect applications. Both stoichiometric as well as nonstoichiometric disorder can lead to alloy scattering, resulting in a strong increase of the resistivity.[20] Several ordered intermetallic aluminides,[21-23] in particular NiAl,[24-26] have been proposed and studied as potential interconnect metals. However, an ultralow resistivity at small film thickness has not been demonstrated for these aluminides yet and the search for additional binary intermetallics continues. Here, we discuss the suitability of the binary aluminide $Al_3Sc$ as an interconnect metal candidate,[23] based on a combination of thin film studies as well as first-principles simulations of its electronic structure. $Al_3Sc$ combines a low bulk resistivity (7 μΩcm)[27] with a high melting point (1320°C),[28] higher than that of Cu (1085°C), which can be used as a proxy for interconnect reliability. However, only few reports have addressed the properties of $Al_3Sc$ thin films,[29,30] with few data available for the thickness range on the order of 20 to 30 nm, which is relevant for the application. In this work, we demonstrate Al$_3$Sc thin films with a resistivity as low as 12.6 μΩcm at 24 nm thickness. We further discuss the effect of stoichiometry as well as surface oxidation on the properties of the films.

**METHODOLOGY**

All Al$_x$Sc$_{1-x}$ films have been deposited in a Canon Anelva physical vapor deposition (PVD) cluster tool on 300 mm Si (100) wafers. Prior to PVD, a 100 nm thick thermal SiO$_2$ was grown to provide electrical insulation. PVD was performed by co-sputtering from Al and Sc targets at room temperature. Post-deposition annealing was performed in H$_2$ for 30 min at atmospheric pressure. Compositions have been derived from flux calibrations using Rutherford backscattering spectrometry (RBS) with an accuracy of about 1%. The compositions were confirmed on thicker stoichiometric films by RBS and wavelength-dispersive x-ray analysis (WDX). Absolute compositions are accurate to 0.5% Al, whereas the precision (i.e. the difference between compositions) is estimated to be better than 0.1%.

The crystal structure and the thicknesses of the films were determined by grazing-incidence x-ray diffraction (GIXRD, ω = 0.3°) and x-ray reflectance (XRR), respectively, using a Bruker JVX7300 diffractometer with Cu Kα radiation. Film thicknesses were measured by XRR before and after annealing and no thickness variation was observed within experimental accuracy. *In-situ* XRD (IS-XRD) measurements during ramp annealing were performed in a custom-built heating stage in a 5%H$_2$/He atmosphere at a heating rate of 0.2 K/s. The heating stage was mounted on a goniometer that was equipped with a VANTEC 2D x-ray detector and allowed for XRD measurements in a 2θ–ω configuration using Cu Kα radiation. This enabled the measurement of the evolution of Al$_x$Sc$_{1-x}$ XRD patterns as a function of annealing temperature *in-situ*. Additional information about crystallinity, thickness, and chemical composition of the films were obtained by transmission electron microscopy (TEM) using a FEI Titan electron microscope operating at 200 kV. The resistivity of the films was determined from the sheet resistance (Rs), measured in a KLA Tencor RS100 system, in combination with the XRR film thickness, corrected for the thickness of the surface oxide, which was assumed to be insulating. Atomic force microscopy (AFM) was performed on a Bruker ICON microscope in tapping mode. The resistivity at cryogenic temperatures was obtained in a Quantum Design Physical Property Measurement System using patterned Hall bars.

The electronic properties of $Al_3Sc$ were determined by first-principles density-functional theory (DFT) simulations. These simulations were performed using the QUANTUM ESPRESSO package, and the valence electron shells of the elements were represented by Garrity–Bennett–Rabe–Vanderbilt (GBRV) pseudopotentials[31,32], with a kinetic cutoff energy of 60 Ry for the truncation of the plane-wave expansion of the wavefunction. The exchange-correlation energy was described within the Perdew-Burke-Ernzerhof generalized gradient approximation[33]. The first Brillouin zone was sampled by a discretized Monkhorst-Pack scheme[34] based on a regular unshifted (Γ-point centered) $k$-point mesh of 40×40×40. This approach ensured a convergence of the total energy within $10^{-12}$ eV.

**RESULTS AND DISCUSSION**

The GIXRD patterns of 24.1 ± 1.1 nm thick as-deposited $Al_xSc_{1-x}$ films as a function of the Al mole fraction $x$ around stoichiometric $Al_3Sc$ ($x$ = 0.75) indicate that all films were x-ray amorphous (not shown). Figures 1a and 1d show the GIXRD patterns of the 24.1 ± 1.1 nm thick $Al_xSc_{1-x}$ films after post-deposition annealing at 500°C and 600°C respectively, in $H_2$. All patterns are consistent with the $L1_2$ crystal structure (space group $Pm\bar{3}m$) of bulk $Al_3Sc$.[27,28] A comparison with the corresponding 2θ-ω XRD pattern (not shown) indicated that the films were polycrystalline with random grain orientation and no signs of texture. We note the presence of a secondary phase for the two most Al-rich films after annealing at 600°C. This will be discussed further below. *In situ* XRD during ramp annealing (Fig. 1b) of a stoichiometric 25.0 nm thick $Al_3Sc$ film revealed weak crystallization (amorphous-to-crystalline transition) at 193°C, followed by recrystallization (crystalline-to-crystalline transition) at 443°C,[35,36] as extracted from the evolution of the (111) peak intensity with annealing temperature (Fig. 1c). The lattice parameter extracted from the GIXRD pattern at 500°C showed a linear decrease with increasing Al content (Fig. 1e), consistent with the relative atomic radii of Al and Sc. For stoichiometric $Al_3Sc$ ($x$ = 0.75), it was consistent with the bulk value of 4.10 Å[27,28] within experimental accuracy.

The crystalline nature of the annealed stoichiometric $Al_3Sc$ films was confirmed by the plan-view transmission electron micrographs in Figs. 1f and 1g, both taken after annealing at 600°C in $H_2$. The images show a coexistence of large grains with sizes above 100 nm with zones of much smaller grains. This hints towards partial recrystallization with zones of immobile grain boundaries that have not yet been recrystallized. [35,36]

Figure 2a shows the composition dependence of the resistivity of 24.1 ± 1.1 nm thick $Al_xSc_{1-x}$ films as a function of their composition. As deposited films showed resistivities on the order of 100 μΩcm with little dependence on composition, which is consistent with an x-ray amorphous random alloy. However, the film (re-)crystallization induced by post-deposition annealing strongly reduced the resistivity with a resistivity minimum appearing for the stoichiometric $Al_3Sc$ ($x$ = 0.75) film. This behavior is reminiscent of that of NiAl[24,26] and can be attributed to a combination of crystallization (*cf.* Fig. 1) and possibly $L1_2$ ordering. It should be mentioned that the degree of $L1_2$ ordering is difficult to extract from the XRD data since the difference with random fcc XRD patterns is very small. Hence, additional measurements are required to establish the contribution of intermetallic (dis-)order to the thin film resistivity.

The minimum resistivity was found for the stoichiometric $Al_3Sc$ film with a resistivity of 12.6 μΩcm after annealing at 500°C and above. Lowest resistivity for ordered stoichiometric intermetallics can be expected, since nonstoichiometric point defects lead to alloy scattering and typically strongly increase the resistance. This can be compared to NiAl, for which the resistivity of much thicker films (10 μΩcm at 260 nm,[24] 13 μΩcm at 56 nm[23,26]) also showed a sharp resistivity minimum for stoichiometric films with comparable values. At thicknesses around 25 nm, the Al$_3$Sc resistivity was however considerably lower than that of NiAl, which showed a resistivity of 20.2 μΩcm for a 25.1 nm film under the same annealing conditions.[26] As discussed above, the sensitivity of the resistivity to finite size effects (at identical grain size and film thickness) is governed by the MFP of the metal. The MFP λ of metals has been determined from the product of the bulk resistivity ρ$_0$ and λ, obtained by DFT within a semiclassical transport framework.[5] The advantage of this approach is that, within the approximations of this framework, ρ$_0$×λ depends on the Fermi surface morphology only without the need for detailed scattering calculations.[5,37] Using a numerical integration scheme applied to the Fermi surface of Al$_3$Sc in Fig. 2b and the procedure detailed in Refs. 5,12, and 37, the calculated ρ$_0$×λ product within the constant-λ approximation[5,37] was 4.9×10$^{-16}$ Ωm$^2$, very similar to that of NiAl (4.4×10$^{-16}$ Ωm$^2$) and Ru (5.1×10$^{-16}$ Ωm$^2$), and considerably lower than that of Cu (6.8×10$^{-16}$ Ωm$^2$).[5,12] Using an experimental value for the bulk resistivity of 7 μΩcm,[27] this leads to a MFP for Al$_3$Sc of about 7 nm, *i.e.*, a value much shorter than the Cu MFP of 40 nm. The above results thus show that Al$_3$Sc can be considered promising for future low resistance scaled interconnect metallization schemes in advanced microelectronic technology nodes, since it combines low bulk resistivity with an ultrashort MFP.

To gain further insight into the transport properties of Al$_3$Sc thin films, we have measured the temperature dependence of their resistivity (Fig. 2c). The resistivity increased nearly linearly with temperature over a large range, which indicates that temperature dependence was mainly determined by phonon scattering. The deviation from linearity around 80 K is not yet understood but may be linked to the phonon structure.[38] The residual resistance ratio, RRR = ρ$_{300K}$/ρ$_{10K}$ was 1.5, much lower than the value of 7 observed for bulk samples.[27] This suggests a strong contribution by temperature-independent grain boundary scattering to the resistivity, which reduces the RRR. In addition, residual disorder (point defects) also leads to temperature-independent scattering and a reduction of the RRR. A separation of the two contributions is however not possible without a detailed knowledge of the nature of the defects present in Al$_3$Sc, their density, and their scattering rates. A rule of thumb previously applied to other intermetallic aluminides[39] suggests that the disorder of the L1$_2$ structure in our films could be on the order of 1%, assuming that the reduction of the RRR is attributed to disorder effects alone. This further suggests that the films are highly ordered.

To further study the structure of the films, we have measured the rms surface roughness of 24.1 ± 1.1 nm thick Al$_x$Sc$_{1-x}$ films by AFM as a function of their composition after annealing at 600°C. The results in Fig 3a show low rms roughnesses below 1 nm, except for the most Al-rich Al$_x$Sc$_{1-x}$ films with $x \geq 0.79$. For these films, the surface roughness is as large as 10 nm, which can be attributed to large-scale dendrite-like structures with heights up to 100 nm (Figs. 3b and 3c). This is concurrent with the observation of a secondary phase in the GIXRD patterns of the two most Al-rich films with $x \geq 0.79$ after 600°C annealing in Fig. 1d. The GIXRD patterns show two peaks/shoulders around 38.5° and 44° that are consistent with fcc Al. The origin of a third peak around 35.5° is currently unknown.

The observation of Al clusters is in agreement with the Al-Sc phase diagram[28] (Fig. 3d) that indicates that Al$_3$Sc is a line compound with negligible miscibility of Al or Sc in the intermetallic. Hence, excess Al (or Sc) should lead to phase separation and secondary phase formation. Our experiments show that the formation of a secondary Al phase (and possibly a third unidentified phase) becomes visible for the studied thermal budgets in thin films in form of large dendrite-like structures around 4% of excess Al after 600°C post-deposition annealing. Nonetheless, it is possible that some limited Al secondary phase, with a concentration below the AFM and GIXRD detection limits, was already present for lower Al excess or at lower annealing temperatures. The presence of a low-resistivity Al secondary phase may also explain the decrease of the resistivity of the most Al-rich Al$_x$Sc$_{1-x}$ films with $x \geq 0.79$ (Fig. 3a) after annealing. By contrast, the expected phase separation in case of Sc excess, *i.e.,* the formation of a Al$_2$Sc secondary phase, was not visible in the data, possibly due to a much smaller length scale for phase separation that renders the expected Al$_2$Sc clusters difficult to detect. We note that this behavior was markedly different from that of NiAl, which has been previously studied for interconnect applications[24-26] and possesses an extended phase field with considerable solubility of excess Ni or Al.

The polycrystalline microstructure of stoichiometric $Al_3Sc$ films after annealing at 600°C was further confirmed by cross-sectional TEM imaging. The micrograph in Fig. 4a is consistent with the plan-view TEM images in Fig. 1f and 1g, showing the coexistence of large and small grains, separated by inclined ($\theta \sim 72°$) bamboo-like grain boundaries. The high-angle annular dark field (HAADF) image in Fig. 4b shows that the composition of the film was uniform.

The images further show the presence of both a ~6 nm thick surface oxide (due to air exposure, both after deposition and annealing) and an about 4 nm thick interfacial layer in contact with the underlying $SiO_2$. Chemical profiling using electron-dispersive spectroscopy (Fig. 4c) indicated that the surface oxide was non-stoichiometric (Sc-rich) with Al/Sc ~ 2, while the ratio was consistent with the expected value of 3 (within experimental accuracy) in the central part of the film. We note that a similar nonstoichiometric surface oxide was already observed for NiAl.[26] Since the nonstoichiometric surface oxide was already present on as-deposited films (data not shown), and can therefore not stem from *e.g.* phase separation during post-deposition annealing, this suggests that the surface oxidation was limited by metal out-diffusion rather than O in-diffusion, with Sc diffusing faster to the surface than Al. We note that the detection of low O levels in the bulk part of the $Al_3Sc$ film may stem from the sample preparation or measurement artifacts, since XPS profiling (not shown) found O levels to be on the order of the background of 1 at.%. Finally, the chemical profile also shows that annealing at 600°C led to a reaction with the underlying $SiO_2$, forming an inhomogeneous $AlScSiO_x$/$AlScO_y$ layer. Thus, the interface formation of $Al_3Sc$ on low-κ dielectrics requires further study and could necessitate the introduction of dielectric barrier layers or the optimization of the thermal budget for interconnect integration.

**CONCLUSION**

In conclusion, we have reported the properties of $Al_xSc_{1-x}$ thin films with compositions around $Al_3Sc$ ($x$ = 0.75), relevant for interconnect metallization applications. As-deposited films were x-ray amorphous and (re-)crystallized above about 450°C, with an associated strong reduction of the resistivity. After annealing at 500°C, 24 nm thick $Al_3Sc$ showed a resistivity of 12.6 μΩcm, considerably lower than previously reported values for NiAl, the so far most studied aluminide intermetallic for interconnect applications.[24-26] The optimum resistivity was obtained for stoichiometric $Al_3Sc$ and an annealing temperature of 500°C or higher. Further thickness scaling will rely on the control of (non-stoichiometric) native surface oxide formation that strongly affects the stoichiometry of ultrathin films. Non-reactive *in-situ*-deposited dielectrics or capping layers can be employed to avoid the complications with composition control at layer thicknesses below 15 nm. Additionally, reducing the thermal budget required to achieve low resistivity films offers a potential solution.

**ACKNOWLEDGEMENTS**

This work was supported by imec's industrial affiliate program on nano-interconnects. The authors would like to thank Patrick Carolan, Hugo Bender, Kris Paulussen, Pieter Lagrain, Olivier Richard, Paola Favia, and Laura Nelissen for the TEM analysis; Inge Vaesen and Thierry Conard for the XPS profile; and the imec p-line for their support.

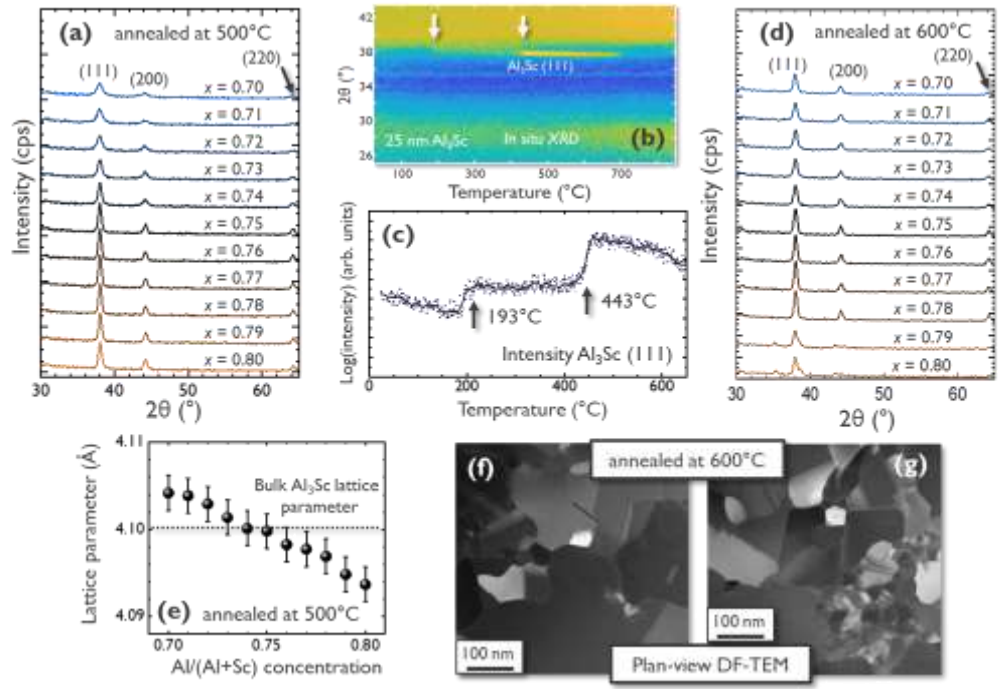

**Figure 1. (a)** GIXRD patterns of 24.1 ± 1.1 nm thick $Al_xSc_{1-x}$ films as a function of Al mole fraction $x$ after post-deposition annealing at 500°C. **(b)** IS-XRD pattern during ramp annealing for a 25.0 nm thick $Al_3Sc$ ($x$ = 0.75) film. The intensity of the $Al_3Sc$ (111) in **(c)** shows crystallization and recrystallization at 193°C and 443°C, respectively. **(d)** GIXRD patterns of 24.1 ± 1.1 nm thick $Al_xSc_{1-x}$ films as a function of Al mole fraction $x$ after post-deposition annealing at 600°C. **(e)** Lattice parameter deduced from (a) as a function of Al mole fraction $x$. The lattice parameter of the $Al_3Sc$ ($x$ = 0.75) film is close to the reported bulk lattice parameter.[27,28] **(f)** and **(g)** show plan-view dark-field (DF) TEM images of a 30 nm thick $Al_3Sc$ film after 600°C annealing.

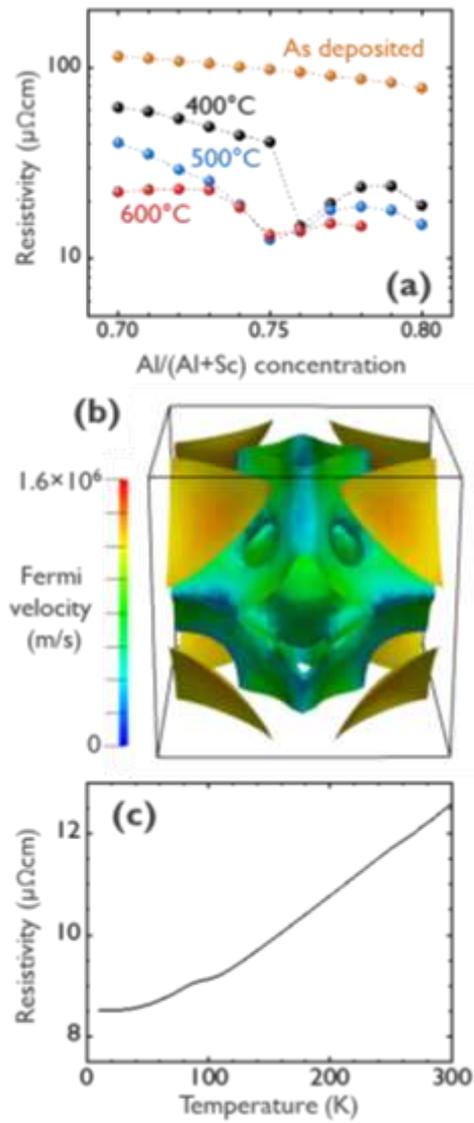

**Figure 2. (a)** Resistivity of 24.1 ± 1.1 nm thick Al$_x$Sc$_{1-x}$ films as a function of Al mole fraction $x$ both as deposited and after post-deposition annealing at the indicated temperatures. **(b)** Calculated Fermi surface of L1$_2$ Al$_3$Sc. The color represents the intensity of the Fermi velocity. **(c)** Resistivity *vs.* temperature of a 25.0 nm thick Al$_3$Sc film after annealing at 500°C.

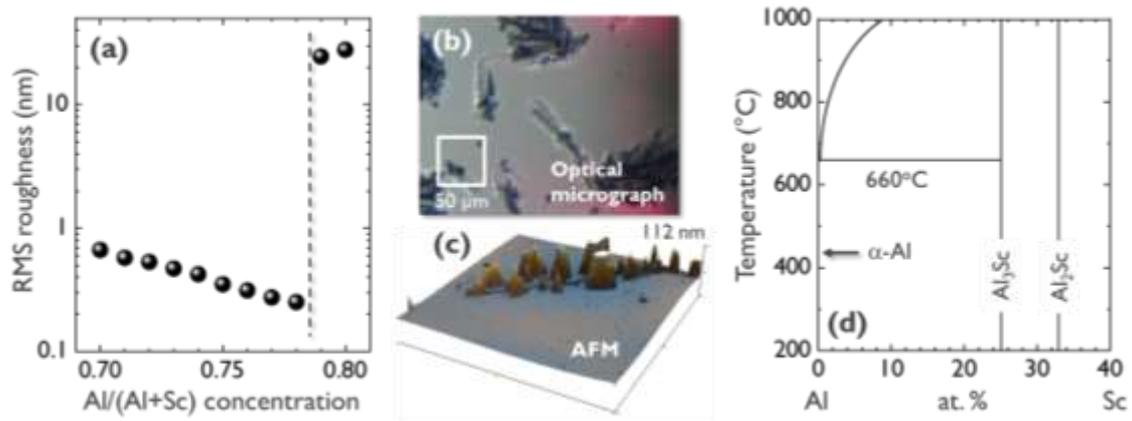

**Figure 3. (a)** RMS roughness of 24.1 ± 1.1 nm thick $Al_xSc_{1-x}$ films as a function of Al mole fraction *x*, deduced from AFM images. **(b)** Optical micrograph and **(c)** 5 μm × 5 μm AFM topography image of the most Al-rich $Al_{0.80}Sc_{0.20}$ film. **(d)** Al-rich section of the Al-Sc phase diagram (redrawn from Ref. 28).

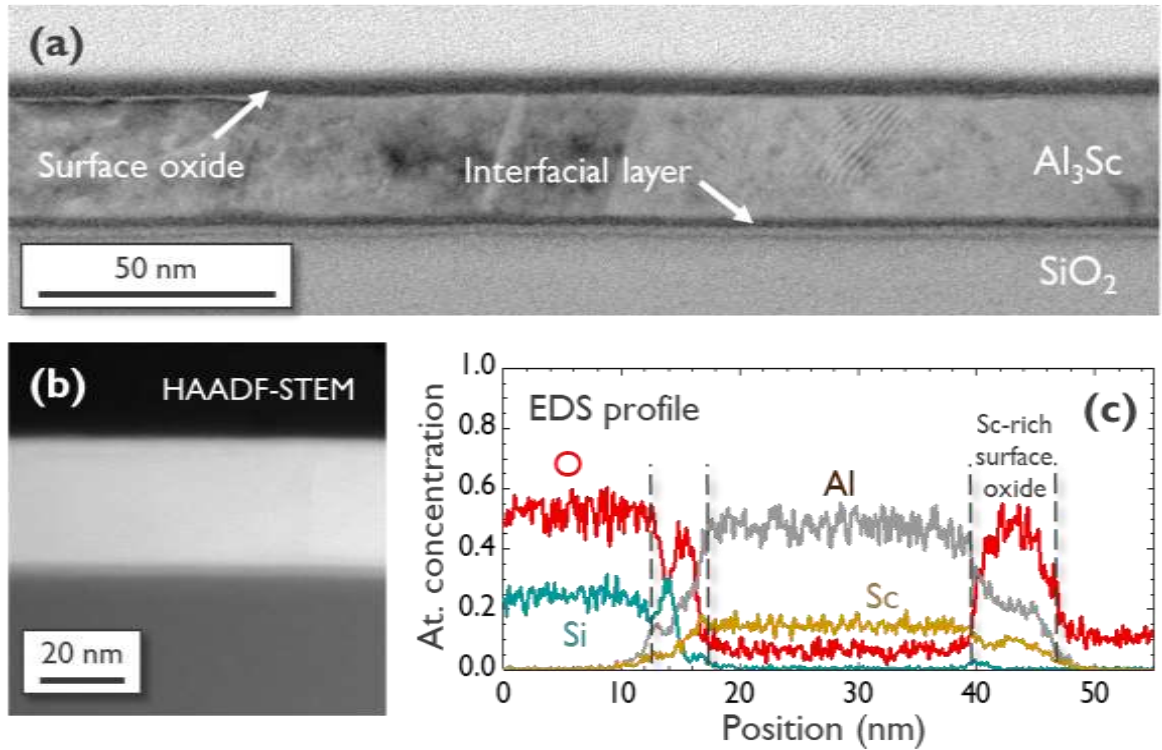

**Figure 4. (a)** Cross-sectional TEM image of an Al$_3$Sc film after annealing at 600°C. **(b)** High-angle annular dark field scanning TEM (HAADF-STEM) image of the Al$_3$Sc film. **(c)** Composition profile across the Al$_3$Sc film obtained by energy-dispersive spectroscopy (EDS).